\documentclass[10pt,conference]{IEEEtran}
\usepackage{cite}
\usepackage{amsmath,amssymb,amsfonts}
\usepackage{graphicx}
\usepackage{textcomp}
\usepackage[hyphens]{url}
\usepackage{fancyhdr}
\usepackage{hyperref}
\usepackage{booktabs, multirow} 
\usepackage{soul}
\usepackage{changepage,threeparttable} 

\usepackage{balance}
\usepackage{graphicx}
\usepackage{listings}
\usepackage{amsfonts}
\usepackage[mathscr]{euscript}
\usepackage{tikz}
\newcommand*\circled[1]{\tikz[baseline=(char.base)]{
            \node[shape=circle,fill,inner sep=1.2pt] (char) {\textcolor{white}{#1}};}}

\definecolor{codegreen}{rgb}{0,0.6,0}
\definecolor{codegray}{rgb}{0.5,0.5,0.5}
\definecolor{codepurple}{rgb}{0.58,0,0.82}
\definecolor{backcolour}{rgb}{0.95,0.95,0.92}

\usepackage{tikz}
\definecolor{olive}{rgb}{0.33, 0.42, 0.18}
\definecolor{keywordcolor}{HTML}{cc33ff}

\usepackage{couriers}
\def\sensitive{0}

\usepackage{pifont}

\usepackage{epsfig,endnotes,enumitem,makecell,textcomp,graphicx,filecontents,caption,subcaption}

\usepackage{algorithm}
\usepackage{algpseudocode}
\usepackage{url}

\newcommand{\LibName}{LibPreemptible }

\usepackage[normalem]{ulem}

\begin{document}

\title{LibPreemptible: Enabling Fast, Adaptive, and Hardware-Assisted User-Space Scheduling} 
\author{Yueying Li, Nikita Lazarev, David Koufaty, Yijun Yin, Andy Anderson, \\ Zhiru Zhang, Edward Suh, Kostis Kaffes, Christina Delimitrou}

\date{}

\thispagestyle{empty}

\maketitle
\begin{abstract}
       Modern cloud applications are prone to high tail latencies since their requests typically follow highly-dispersive distributions. Prior work has proposed both OS- and system-level solutions to reduce tail latencies for microsecond-scale workloads through better scheduling. Unfortunately, existing approaches like customized dataplane OSes, require significant OS changes, experience scalability limitations, or do not reach the full performance capabilities hardware offers. 	
       We propose LibPreemptible, a preemptive user-level threading library that is flexible, lightweight, and scalable. LibPreemptible is based on three key techniques: 1) a fast and lightweight hardware mechanism for delivery of timed interrupts, 2) a general-purpose user-level scheduling interface, and 3) an API for users to express adaptive scheduling policies tailored to the needs of their applications. Compared to the prior state-of-the-art scheduling system Shinjuku, our system achieves significant tail latency and throughput improvements for various workloads without the need to modify the kernel. We also demonstrate the flexibility of LibPreemptible across scheduling policies for real applications experiencing varying load levels and characteristics.
\end{abstract}
\section{Introduction}

A large portion of the world's computation is now hosted on either public or private cloud infrastructures. This has brought on several changes to the way cloud applications are designed, including
moving away from monolithic services to inter-dependent microservices and event-driven serverless frameworks~\cite{Barroso17,Gan19,seer,sage,utune,siw}. 
These software design approaches 
enable high concurrency of fine-grained tasks, with thousands of user requests executing at any point in time. The fine-grained nature of these tasks also allows them to be coscheduled on the same physical host, which facilitates multi-tenancy and improves the cloud's resource efficiency. At the same time, applications must meet service-level objectives (SLOs), often defined in terms of tail latency.

Furthermore, datacenter suffers from \textit{massive thread over-subscription}. We observe serious thread over-subscription consistently from recent Google traces~\cite{googleworkloadtraces} across widely-used applications, where more than 50 threads on average or sometimes around 500 threads can be scheduled to each core (Table~\ref{tab:oversubscription}). To achieve higher CPU efficiency and low tail latency, one necessary precondition is fine-grained, scalable, adaptive, and low-overhead preemptive scheduling.

Preemption enables more fine-grained resource sharing among workloads and requests. Without preemptions, short requests can get stuck behind long requests, causing head-of-line (HoL) blocking. The lack of fine-grained preemptions results in prior work experiencing high latencies under long-tailed request service time~\cite{Prekas2017,Daglis2019,IX,Lim2014,Demoulin2021}. 

\begin{table}
    \footnotesize
    \centering
    \begin{tabular}{|c|c|c|c|}
    \hline App (code name) & \# of threads & \# of cores & Threads/core \\
    \hline charlie & 4842 & 10 & 484 \\
    \hline delta & 300 & 4 & 75 \\
    \hline merced & 5470 & 110 & 50 \\
    \hline whiskey & 1352 & 8 & 169 \\
    \hline
    \end{tabular}
    \caption{Datacenter thread oversubscription from four widely used applications in Google~\cite{googleworkloadtraces}.}\label{tab:oversubscription}
    \vspace*{-0.30in}
\end{table}

Unfortunately, past proposals for preemptive scheduling are not suitable for microsecond-scale workloads running on exisiting cloud platform, for the following reasons. 1) Preemptive user level threads based on regular interrupts still incur high overheads during user- and kernel- level context switches~\cite{Goproposal,libinger}; if the minimum time slice is 5ms and there are 200 threads on average per core, the scheduler cycle will be increased to 1 seconds, significantly increasing tail latency.  2) Optimizing preemption overhead, like Shinjuku~\cite{Shinjuku}'s usage of Posted IPI can reduce the time slice; however, assigning the programmable interrupt controllers (APICs) to their runtime introduces security concerns for a shared cloud environment. Similarly, using APICs limits the approach's scalability, as it only supports a small number of logical cores.

An alternative scheduling approach to achieving low latency scheduling without preemption is by using request-specific knowledge. 
However, such information is difficult to obtain beforehand.  Additionally, the duration of these requests' execution time has no upper bounds. This makes it challenging to apply rules, such as SRPT (Shortest Remaining Processing Time \cite{harchol2000implementation}) or any other rules that give priority to short requests, especially when operating at the microsecond scale~\cite{Delimitrou13}. 

We propose \textit{LibPreemptible}, a hardware-assisted user-level threading library that enables fine-grained, configurable, and scalable preemptions in the cloud. LibPreemptible is built on top of user interrupt (UINTR), a new hardware capability in the Intel Xeon Scalable Processor codenamed Sapphire Rapids (SPR). UINTR is a low-overhead communication mechanism that allows application threads to directly send each other interrupts bypassing the kernel. However, native UINTR is not a good fit for dynamic workloads without knowledge of request service times. We leverage UINTR to design a user-level threading library that enables configurable and fine-grained periodic interrupts, and adjusts scheduling timeslices on the fly using statistical tests. LibPreemptible also introduces fast and accurate user timers, which can be used by a wide range of applications to easily build customizable scheduling policies. 

We evaluate LibPreemptible via microbenchmarks, as well as synthetic and real applications widely used in private and public clouds today. We show that it achieves better performance, scalability, isolation, and flexibility compared to prior work. The tail latency with LibPreemptible is 10$\times$ better compared to Shinjuku~\cite{Shinjuku}, the previous state-of-the-art system, across various workloads. 
In an environment where applications time-share CPU cores, where a latency-critical application shares resources with a best-effort (BE) compression workload, LibPreemptible achieves on average 10\% higher throughput for the BE job, while
maintaining the same 99\% tail latency SLO for the latency-critical application. %

\textcolor{black}{LibPreemptible's main contributions are: 
\begin{itemize}[leftmargin=*]
    \item First paper to leverage a new hardware mechanism (UINTR) for fine-grained user-level interrupts, that scales much better than previous solutions and faster than traditional interrupts' millisecond timescales~\cite{libinger}. 
    \item Designing a new abstraction with user-level timers with fine-grained deadlines, called LibUtimer, accounting for the diverse needs of different microsecond-scale applications with a wide spectrum of scheduling policies.
    \item Running entirely outside the kernel and only requiring a kernel modification to regularly enable UINTR~\cite{UINTR-patch}. Applications using LibPreemptible can coexist with traditional applications and common cloud system software / hardware stacks with only a few lines of code change.
\end{itemize}
}

This paper focuses not only on just putting UINTR into use, but more on how we should build new abstractions and delegate scheduling decisions to user-level applications, offering adaptive policies that best meet their needs.

\begin{figure}
    \centering
    \begin{tabular}{cc}
        \hspace*{-0.24in}
        \raisebox{0.11\height}{\includegraphics[width=0.48\columnwidth]{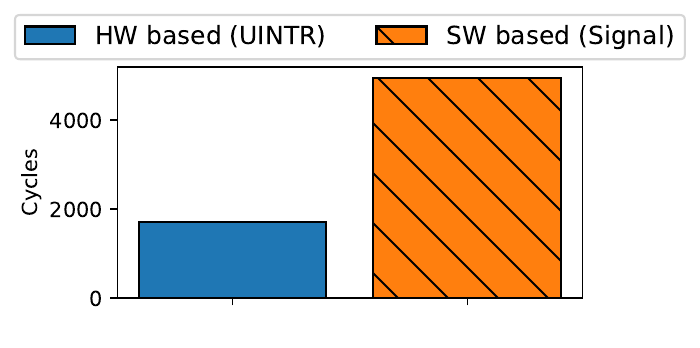}}
        \hspace{-0.03\columnwidth} 
        \includegraphics[width=0.48\columnwidth]{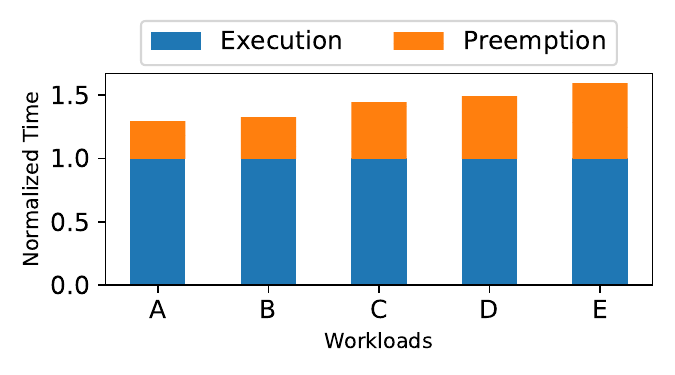}
        \hspace*{-0.14in}
    \end{tabular}
    \caption{Left: Performance gap between software- and hardware-based IPC delivery. Right: Normalized overhead of preemption w.r.t lean execution time for different microsecond-scale workloads running on Shinjuku (ranked by workload dispersion).}
    \label{fig:ipi1}
\end{figure}

\section{Motivation}
\label{sec:background}
Below we demonstrate the characteristics of cloud workloads that motivate the need for a preemptive user-level threading library that operates adaptively, at microsecond scale. 
\subsection{The Need for a $\mu s$-scale Preemption Mechanism}
Previous work has demonstrated that preemption in user space is key to achieve low tail latency\cite{Ousterhout2019,Shinjuku}. Synchronous mechanism offers relatively low latency, but are not practical as they require the worker thread to be idle (for example, blocked in the kernel waiting for an eventfd, or polling or mwaiting on shared memory). Preemption requires the ability to asynchronously receive an event during execution of a user-level thread to trigger the context switch. 

There are two approaches to deliver asynchronous events to drive preemptions; however, the overhead, lack of precision, modifications and security concerns often overshadow the benefit on performance. 

One approach to delivering the asynchronous event is to use signals~\cite{libinger, Goproposal, Shiina2021}. Typically, a thread creates a kernel timer at a specific future time and the kernel delivers a signal to the thread when the timer expires. In some cases, threads additionally communicate using signals to propagate the preemption event to other threads. However, a kernel timer is susceptible to kernel jittering and the signal overhead is significant and scales poorly as the number of threads increases due to signal contention (Section~\ref{sec:eval_analysis}). These reasons limit the applicability of this approach, and result in relatively large preemption timers, as in the case of the Go programming language~\cite{Goproposal}, which recently introduced preemption to prevent starvation at a 10ms granularity.

Another approach is to use native interrupts, which provide a low latency mechanism for inter-processor communication (IPC). However, interrupts are restricted to kernel software due to its privileged nature, so it's challenging to use them directly within application threads. Furthermore, using interrupts would require system calls at sender threads and signals at the receiver threads. Prior works, such as Shinjuku~\cite{Shinjuku}, have circumvented this overhead at the sender by mapping the physical APIC to the sender, which is untenable in the public cloud environment. Receiving events in user space still requires kernel-mediation (for example, to notify workers threads of the interrupt using signals)~\footnote{Other interrupt optimizations explored in Shinjuku (to avoid VM exits at sender) in a virtualized environment are becoming irrelevant as hardware vendors introduce support for inter-processor interrupts (IPI) virtualization~\cite{ipi_virtual}.}. 

We profiled the overall CPU time spent in preemption vs. execution (normalized to execution time) with the time quantum chosen to offer the best tail latency for various latency-critical workloads running on baseline system~\cite{Shinjuku}. It can be observed that preemption overhead is significant, especially for micro-second scale workloads with high dispersion (Figure~\ref{fig:ipi1} Right). 
Figure~\ref{fig:ipi1} Left shows the gap between software-based IPC using signals, and hardware-assisted IPC using UINTR. 
While previous software based optimizations achieve performance benefits compared to regular interrupts, there is still a gap to the latency of delivering and handling an interrupt with hardware.

\begin{figure}%
  \centering
	\vspace{0.08in}
	\begin{tabular}{cc}
		\vspace*{-0.05in}
    \includegraphics[width=0.48\columnwidth]{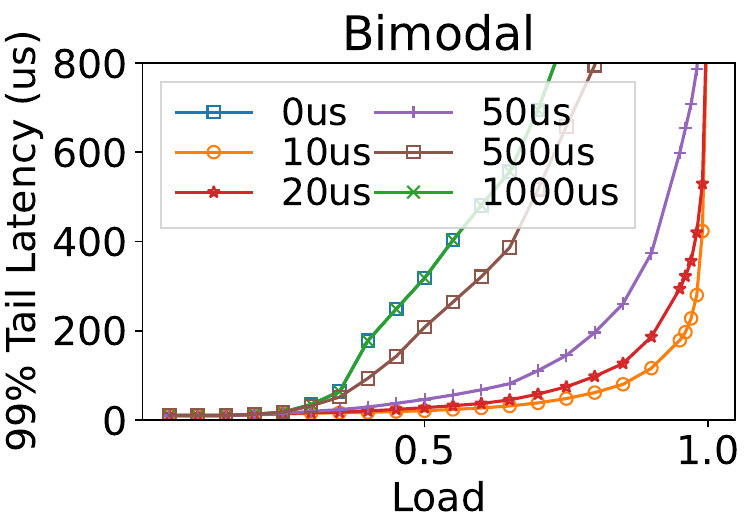}
    \includegraphics[width=0.48\columnwidth]{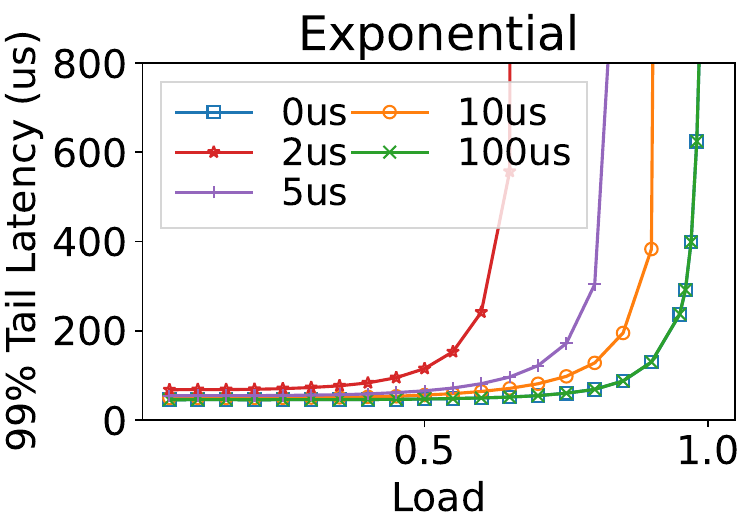}
	\end{tabular}
  \vspace{-0.10in}
	\caption{Tail latency with different preemption quanta on bimodal (left) and exponential (right) workloads. Bimodal: 99.5\% of requests are short (10us) and 0.5\% are long (1000us). Exponential: mean as 10us. The preemption overhead, including sender and receiver cost, is about 1us across all settings.}%
  \label{fig:sens}
\end{figure}


\subsection{The Need for Adaptive Preemption}

Datacenter workloads have a wide variation in request processing times, especially at the microsecond level. Factors such as high load, imbalanced requests, low CPU processing speed due to interference, saturation, rate limiting, etc as well as high 
cache miss ratios due to skewed key distributions, and more, can cause high tail latencies in systems like Memcached~\cite{memcached}. 

Imbalanced request times can lead to head-of-line (HoL) blocking where short requests are blocked by longer ones, making adaptive scheduling policies necessary. The tail-optimal scheduling policy changes with the request service time distribution. To arrive at the best policy, we need to adapt to the workload. For example, in heavy-tailed workloads, c-FCFS (centralized first come first serve) scheduler with preemption is better than PS (processor sharing), while the throughput depends on the length of the time quantum. A time quantum that is too long causes HoL (Head of Line) issues, while one that is too short results in a decrease in CPU efficiency.

Figure~\ref{fig:sens} illustrates the importance of adaptive preemptive scheduling policies. 
With two typical service time distributions running on 16 cores and different preemption quanta\footnote{ 0us time quantum means no preemption. }, lower preemption quanta give better tail latency in heavy-tailed workloads (e.g. bimodal, until the time quantum becomes too small), wheras higher time quanta give better tail-latency for light tailed workloads such as exponential workloads. The tail-optimal scheduling policy varies with the request service time distribution~\cite{Wierman2012}.

While aggressive preemptive scheduling can help reduce tail latency in highly dispersive workloads, it's not always necessary under lower loads or lighter tailed workloads.  

Current scheduling primitives (including OS or user-level green threads) do not expose such decisions to users for the following fundamental reasons. First, due to semantic mismatch, user applications cannot dynamically adapt CPU resources. Second, user applications can delegate scheduling decisions to kernel, which are coarse grain and not scalable to emerging application requirements.

\section{User Interrupt, Limitation and Design}

We first discuss the background of user interrupt, the challenges of using them, and design objectives of LibPreemptible, then present how we arrive at the right abstraction for fine-grained user-level interrupt delivery which the library is built upon, 
and finally show a framework for building application-specific adaptive request schedulers based on LibPreemptible.

\subsection{User Interrupts}

\pagestyle{plain}

The kernel is used in almost all instances of communication across privilege boundaries, and it includes notification mechanisms based on hardware interrupts, signals, pipes, files, etc.
The only exception is when applications use memory for communication, which requires the receiver to poll, resulting in resource inefficiency. 

User interrupt is a new hardware capability that delivers low overhead preemption without kernel mediation. User interrupts avoid transitions through the kernel, by enabling the possibility of sending and handling hardware interrupts directly in non-privileged applications in the user space~\cite{UINTR}. While their setup requires kernel enabling, in runtime, delivery of a user interrupt to a running thread results in nearly the same latency as that of a regular interrupt. User interrupt also works with inter-processor interrupt virtualization, which allows guest software to send and receive UINTR without the cost of a VM Exit. 

\begin{figure}
  \centering
  \vspace{0.14in}
  \includegraphics[width=0.93\columnwidth]{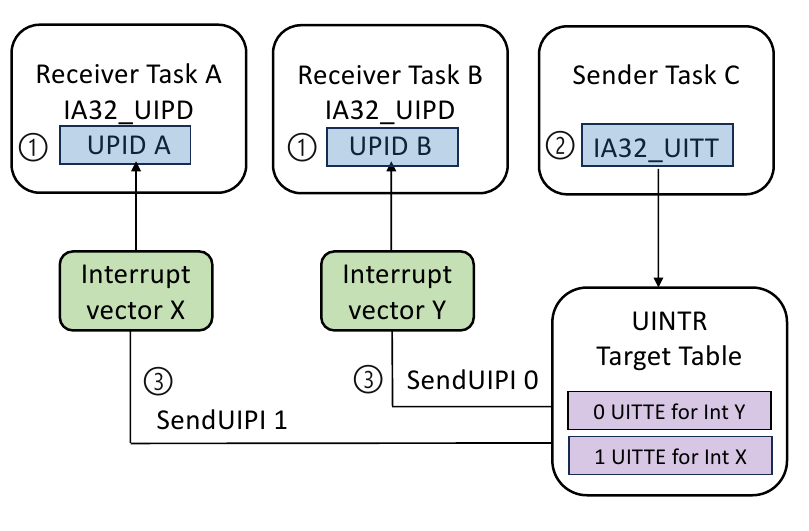}
	\caption{User Interrupt overview. The mechanism can be decomposed into two phases, setup phase and delivery phase.  In the setup phase, 1) receivers set up \texttt{UPID} and register the handler, and return \texttt{uintr\_fd} to the sender; 2) senders use \texttt{uintr\_fd} to allocate \texttt{UITTE} and get the UIPI index. In the delivery phase, 3) senders call \texttt{SENDUIPI}.}%
  \label{fig:uintr}
\end{figure}

The interrupt delivery happens in three steps (Figure \ref{fig:uintr}):  1) Each receiver stores interrupt information in a memory location defined by User Posted Interrupt Descriptor (UPID). 2) Each sender requires a table targeting receivers. The per-thread User Interrupt Target Table (UITT) contains the target UPID and vector to deliver. 
User interrupts have 64 interrupt vectors per thread. 3) SENDUIPI takes an index to the UITT table to deliver an interrupt, and the UPID holds all the information necessary to post the request and send a notification to the CPU. 
 The CPU control flow can be modified by a user interrupt, also called user interrupt delivery. It can be triggered by either the processor state or the state of memory data structures, managed by the processor and OS. 

 If the receiver is runnable, the UPID records the request and suppresses the notification. If blocked, the UPID uses an ordinary interrupt to unblock the receiver and inject the user interrupt. 

\subsection{Challenges and Abstractions for Preemption}
To fully realize the benefit of hardware support for user-level interrupts, we design LibPreemptible, a scheduling library that achieves programming flexibility and dynamic time sharing built on top of UINTR. Native User interrupt (including unmodified UINTR instruction set architecture, and kernel APIs, shown in Figure~\ref{fig:traditional_api})  is not a good fit for micro-second scale scheduling for modern datacenter applications.

\begin{itemize}[leftmargin=*]
  \item First, the lack of exposure to application-level info makes it hard to dynamically change interrupt decisions based on runtime application characteristics. This requires defining an appropriate interface for users to express application requirements.
  \item Second, UINTR employs the security model of eventFD, which may cause DoS across untrusted processes. This requires defining an appropriate protocol to communicate scheduling decisions across untrusting processes without violating their respective constraints.
\end{itemize}

\begin{figure}[t]
  \lstset { %
  language=C++,
  backgroundcolor=\color{black!5}, %
  basicstyle=\footnotesize,%
}

\lstdefinestyle{customcpp}{
aboveskip=0.1in,
belowskip=0.1in,
abovecaptionskip=0.08in,
belowcaptionskip=0in,
captionpos=b,
xleftmargin=\parindent,
language=c++,
showstringspaces=false,
basicstyle={\linespread{0.9}\fontseries{sb}\footnotesize\ttfamily},
keywordstyle=\bfseries\color{keywordcolor},
commentstyle={\itshape\color{green!40!black}\fontsize{8}{8}\selectfont},
backgroundcolor=\color{white},
frame=single
}

	\begin{lstlisting}[style=customcpp]
// Sender API
// Receive FD via inheritance or UNIX domain sockets
uipi_handle = uintr_register_sender(uintr_fd,flags);
int uintr_unregister_sender(uintr_fd, flags);
// x86 Instruction
void _senduipi(uipi_handle);
// Receiver API
// Register User Interrupt Handler
int uintr_register_handler(handler_func, flags);
int uintr_unregister_handler(flags);
// Create an fd representing the vector - priority
uintr_fd = uintr_create_fd(vector, flags);
void __attribute__ ((interrupt)) 
    u_handler(struct __uintr_frame
    *frame, unsigned long vector) {
    write(STDOUT_FILENO, "User Interrupt!\n", 16);
    uintr_received = 1;
}
\end{lstlisting}
\centering
\footnotesize
\caption{Native UINTR API.} 
\label{fig:traditional_api}
\end{figure}

Based on these challenges, we discuss how to arrive at the right abstraction. Realizing fine-grained and adaptive preemption is challenging. OS threads are the natural choice due to the OS support for preemption and a rich set of functionality: per-thread signal masks, CPU affinity, etc. However, their benefits are limited when optimizing tail latency due to multiple reasons, most notably \circled{1} granularity of kernel context switching (in $ms$-scale), \circled{2} overhead of the context switch and \circled{3} limited information about an application's characteristics that the OS can use towards scheduling.

 We thus build a user-level library that can run microsecond-level requests with accurate and fine-grained deadlines as a protocol, down to 3us. This deadline abstraction is hard to achieve due to its fine granularity, but is 
necessary for the following reasons: First, it allows the scheduler to translate latency requirements to deadlines for each requests, preempting long requests when necessary and canceling some requests to release resources when they have violated SLO. Second, it enables the user-level runtime to record past request information in a generic manner. Third, deadlines can be application-specific, and the scheduler adjusts them to different loads and QoS constraints (Sec~\ref{sec:design-scheduler}). Fourth, untrusted applications cannot preempt other applications based on how their timers are compiled (Sec~\ref{sec:design-timer}).


Hence, we propose using a controllable user-level timer (LibUtimer) for applications within the same security domain, along with a lightweight API and runtime support (LibPreemptible) for flexible scheduling with user-level threads. These solutions target the challenges posed by dynamic, unknown workloads. Due to the applications' dynamic nature (both in terms of request service time and various load), an online algorithm is needed to achieve near-tail-optimal scheduling.

\subsection{Design Goals}

We aim to design LibPreemptible, as a general purpose user-level threading library, focusing on four main aspects: speed and fine-granularity of scheduling decisions, scalability, flexibility, and portability. 
In particular, LibPreemptible is designed to meet the following requirements:

\begin{itemize}[leftmargin=*]
	\item {\bf \textit{Fine-granular lightweight preemption:}} the library should provide efficient task preemption mechanisms to allow building schedulers with time quanta in the order of microseconds;
	\item {\bf \textit{Scalability:}} the overhead of the preemption mechanisms should be small, and it should grow sub-linearly with the number of threads in the application;
	\item {\bf \textit{Flexibility:}} the design of LibPreemptible should be decoupled from the design of the scheduling policies, while exposing the minimal API that allows 
		application developers to incorporate their own policies on top;
  \item {\bf \textit{Portability:}} LibPreemptible should be implemented as a standalone user-space library, running on commodity servers, and be compatible with common cloud system software stacks.
\end{itemize}

\subsection{Design Overview}

\begin{figure}
  \centering
  \includegraphics[width=0.94\columnwidth]{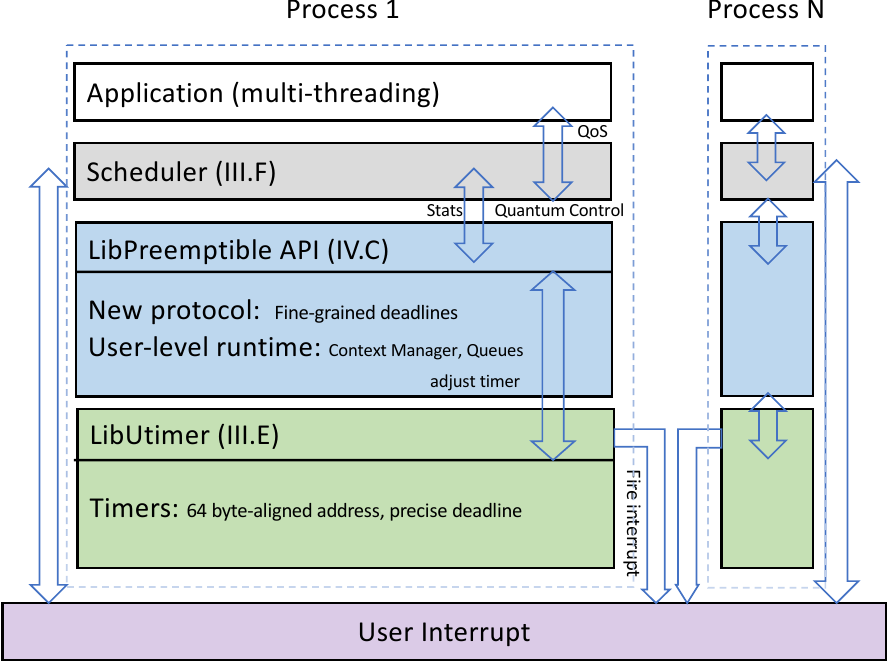}
	\caption{LibPreemptible can be decomposed into three components: Scheduler, LibPreemptible API, and LibUtimer. } 
  \label{fig:structure}
\end{figure}


LibPreemptible provides an API and runtime support for application developers to design efficient latency-critical request processing systems with custom scheduling policies. The library provides interfaces and APIs that are generalizable across application requirements. Figure~\ref{fig:structure} shows how different components interact with each other and with UINTR. LibPreemptible takes the scheduler’s decisions on time quantum to adjust the deadlines (Quantum Control). The queues from LibPreemptible provide statistics of past requests (Stats). LibUtimer adjust timers and fire user interrupt to preempt worker threads. 

\begin{figure}
  \centering
  \includegraphics[width=0.98\columnwidth]{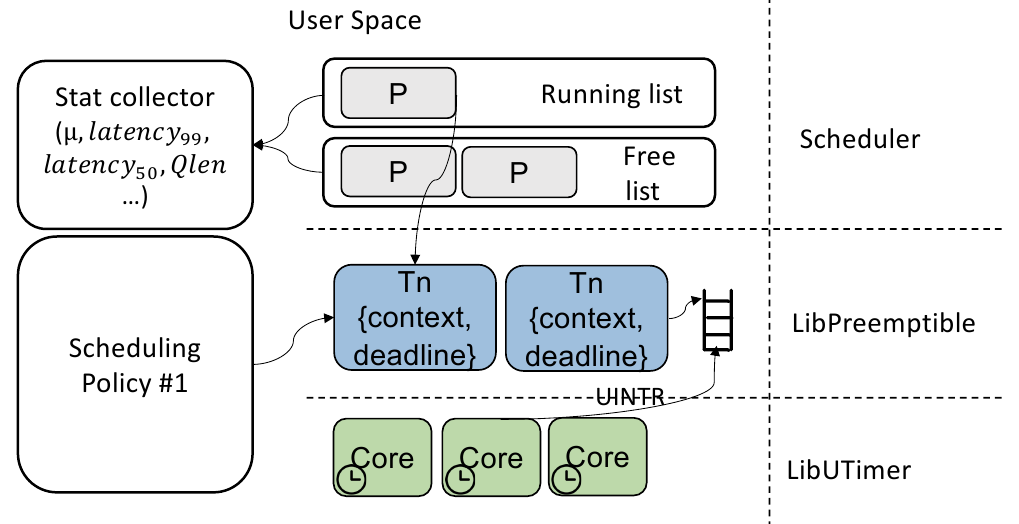}
	\caption{A logical view of an adaptive request scheduler built upon LibPreemptible.}%
  \label{fig:scheduler}
\end{figure}

In particular, Figure~\ref{fig:scheduler} represents how the data structures of LibPreemptible interact with each other. In the scheduler layer, for each request, LibPreemptible defines Function thread (Tn) consisting of the request Context (C) and the Deadline (D). The context encapsulates each request's state (e.g., stack, instruction pointer, etc.) at any point in time. 
In our current implementation, the context is based on the lightweight \texttt{fcontext}. The deadline represents the total time slice given for the requests by the scheduler according to the current policy. When the deadline is reached, LibPreemptible preempts the request by storing the context in the preempted list (also defined as a part of the application) and returning control to the scheduler for the next scheduling decision.

LibPreemptible is based on another library, called LibUtimer. LibUtimer implements an accurate user-space timer that periodically fires preemption signals to all active requests when the corresponding deadlines are reached. The process of delivering these signals is crucial to the performance, scalability, and portability of the library.

\subsection{Preemption Timer} \label{sec:design-timer}

We use UINTR to build an efficient, fast, and scalable preemption timer library LibUtimer.

A simple solution for preemption is to use kernel timers. Kernel timer comes with significant performance costs, mostly due to the kernel overheads of setting up the timer using system calls, and the signal delivery. System call overheads can be reduced by making a single system call with a periodic timer when the preemption interval is static. However, this approach does not work when the timer is dynamic or must be adjusted due to yields. Signals have a significant performance overhead that worsens as the contention in the kernel increases due to timer alignment; for example, when timers are created right after thread creation (creation-time). Prior work~\cite{Shiina2021} tried to reduce this overhead by spreading the timers along the timer interval (staggered) or via a single kernel timer and signals to communicate the timer event to other threads (chaining). 

However, these approaches are impractical for the fine-grained and dynamic timers needed to reduce tail latency, as we will show in Section~\ref{sec:eval_analysis}.  They show poor scalability, as there is no sufficient spacing between timer events to reduce kernel contention, as the thread count increases~\cite{Shiina2021}. The overhead of these timers would offset the benefit of the scheduler in terms of tail latency.

These shortcomings motivate a dedicated user-timer library based on UINTR that scales to a large number of applications issuing concurrent interrupts, reducing context switching overhead, and enabling latency-critical applications to meet their SLOs. LibUtimer leverages timing wheel technique and is able to achieve 3us minimum time slice for scheduling, much lower than any other kernel timers. This fine-grained time unit enables a new design dimension for preemptive online scheduling. The API is described in detail in Section~\ref{sec:LibUtimer}. 

\subsection{User-Level Scheduler} \label{sec:design-scheduler}

LibPreemptible exposes an API for users to easily integrate application-specific scheduling policies.

The scheduler is responsible for scheduling incoming requests across available worker threads. The scheduler is triggered either by the preemption signal from the UINTR, which is controlled by the deadline in the LibPreemptible library or when a new request arrives. It then makes the next scheduling decision based on the set of metrics (Stats) collected from the previous requests over a given time window, typically 10s (including the request load $\mu$, median and tail latencies, the length of the local queues $Qlen$). The statistics are used to control QoS by changing the deadline or time quantum for each request individually. 

In this example, we use a two-level scheduling mechanism to achieve adaptive QoS control with scheduling (Figure \ref{fig:scheduler}). Each worker thread is associated with a local queue of requests to be executed. The local scheduler acts in the following situations: 1) when it sees an incoming request on the queue, 2) or when a function got timed-out/preempted. Local queues for each worker manage execution of functions in a FIFO order. Preempted long-running functions go into the global ``running list'' together with their context. Finished functions go into the global ``free list'' and the contexts are reused. A centralized running and free context list, which the scheduler maintains, help with better load balancing. 

Furthermore, we can adaptively control the time quantum based on past latency distributions and real-time load. Algorithm 1 shows a time quantum controller that manages the change of user-level scheduler dynamically\footnote{For heavy tail identification and definition of tail index on line 7 in the algorithm, please refer to the methodology~\cite{tools00p2, crovella1999estimating}. We use a similar statistical test to determine the direction of changing quanta, that can easily be implemented and enforced with our API under dynamic workloads. A feedback controller can also be integrated with our API. } (which is demonstrated in section~\ref{sec:coloc}). During high load, the preemption interval becomes lower, ensuring timely and precise interrupt delivery to the right application thread. We set the period for the routine of changing time quanta as 10 sec, $L_{high}$ as 90\% of max load, and $L_{low}$ as 10\% of max load. The tail index (0  $\le \alpha < 2$) is considered as a heavy tail distribution~\cite{crovella1999estimating}. The UINTR and LibUtimer enable a minimum time quantum of 3 microseconds, a level that previous mechanisms couldn't attain due to issues with jitter. It ensures microsecond-scale tail latency. In practice, the hyperparameters can be adapted to different workloads from tracking past request data with our API.

\begin{algorithm}[h]
    \footnotesize
    \label{code:controller1}
    \caption{Adaptive Time Quantum Controller}
    \begin{algorithmic}[1]
    \State {\bf Input:} past [$Qlen$, median and tail latencies], incoming load ($\mu$)
    \State {\bf Hyperparameters:} $L_{high}, L_{low}, k_1, k_2, k_3, Q_{threshold}$
    \State {\bf Hyperparameters:} $T_{min}, T_{max}$
    \State {\bf Function:} f (fitting tail index with past statistics) 
    \State $TQ \gets$ current time quantum
    \State  Tail index ($\alpha)\ \gets$ f(past [median and tail latencies])
    \If {$\mu > L_{high}$}
        \State $TQ \gets \min \{TQ - k_1, T_{min}\} $
    \EndIf
\If {[$Qlen > Q_{threshold}$ or tail index ($\alpha$) falls in heavy tail distribution] }
        \State $TQ \gets \min \{TQ - k_2, T_{min}\}$
    \EndIf
    \If {$\mu < L_{low}$}
        \State $TQ \gets \max \{TQ + k_3, T_{max}\} $
    \EndIf
    \State {\bf Output:} updated time quantum ($TQ$)
    \end{algorithmic}
\end{algorithm}

\section{Implementation}
\label{sec:impl}

LibPreemptible is implemented in C, and can be deployed on existing applications with an additional dozen of lines of code  (including setting deadlines when launching user level threads, arming / disarming timers, etc.), and compiled with a standard toolchain, such as \texttt{gcc}. 

\subsection{LibUtimer} 
\label{sec:LibUtimer}

LibUtimer implements fast, hardware-assisted preemptive timers in user space. The user interrupt delivers the preemption notification that will be used 
to trigger scheduling operations if needed.
To deliver the user interrupt, LibUtimer requires that each thread registers a memory location where the time of its next preemption interrupt is located. We refer to this as the \textit{deadline address}, and it is required to be allocated in a dedicated, naturally aligned 64 byte location to avoid false sharing. The deadline specifies the value of the timestamp counter (TSC) when the thread wants its next preemption interrupt. LibUtimer polls on the TSC, and sends a user interrupt to a thread when the TSC reaches its deadline. 

The key interfaces exposed by LibUtimer are as follow:
\begin{enumerate}[leftmargin=*]
    \item \texttt{utimer\_init}: This function creates a pool of timer threads, normally a single thread. This thread will use the RDTSC instruction to check the current value of the TSC and eventually send a user interrupt using the SENDUIPI instruction. It does so by periodically inspecting the value of the TSC and comparing it to the registered deadlines.
    \item \texttt{utimer\_register}: This function is called by application threads to specify the memory address of the deadline. It hides the low-level interactions with the kernel~\cite{UINTR-patch} to register the interrupt handler, create a file descriptor in the context of the application thread, etc.
    \item \texttt{utimer\_arm\_deadline}: This function performs a memory write to set the deadline with the new time to fire the next preemption interrupt. 
\end{enumerate}

A timed interrupt fires periodically, causing our signal handler to be invoked, which in turn fires the user interrupt. We use \texttt{UMWAIT} to free CPUs to go into a low-power state or switch to a hyperthreaded sibling during this short period~\cite{IX, UMWAIT}. Furthermore, for application with large thread counts and request for higher number of timers, we can opt in and use timing wheel techniques~\cite{varghese1987hashed}.

\subsection{Context Management}
To reduce the context switch overhead, we customize the fcontext library~\cite{boost}. The context structure consists of a machine-specific representation of the saved state, the signal mask, a pointer to the context stack, and a pointer to the context that will be resumed when this context finishes execution. The dispatcher allocates context objects and stack space for each request from a global memory pool; an application can define the size of this pool. When scheduler launches a function, a relevant context is attached to it. It is freed when the function related with the context completes execution and is returned to the pool of global contexts. When a function gets preempted, the context will be put into a global wait list. 

Context can be reused by other requests once a function finished execution. The free contexts are maintained in a global free list. We further optimize the context switching overhead as done in Shinjuku~\cite{Shinjuku}.

\subsection{Adaptive User-controlled API}

LibPreemtible exposes a simple API that allows the creation and immediate execution of a function until the function completes, or its time slice has been reached. In either case, control is returned to the caller, which then can decide which function to resume.

The key interface exposed by LibPreemptible is as follow:
\begin{enumerate}[leftmargin=*]
\item \texttt{fn\_launch}: This function is used to create a preemtible function. Its execution begins immediately, and control is returned to the caller when the function completes or a timeout (the time slice) is reached. State for the preemptible function is allocated by the caller, and saved upon preemption.
\item \texttt{fn\_resume}: This function resumes execution of a preemptible function. As with \texttt{fn\_launch}, control is returned to the caller when the function completes or a timeout is reached.
\item \texttt{fn\_completed}: This function checks the status of a preemptible function and indicates its completion before the timeout expired, so that a reschedule is unnecessary.
\end{enumerate}

When requests are scheduled, each runs on top of a lightweight preemptible function. When the request exceeds its timeout, its state is saved, 
including the registers and PC, and control returns to the worker thread. The worker thread can implement a custom scheduling policy, incorporating factors, 
such as the request's wait and execution time, and account for an application's SLO. Figure~\ref{fig:preemption_example} shows a simple example of a round-robin scheduler using the LibPreemptible API.

\begin{figure}[t]
  \lstset { %
  language=C++,
  backgroundcolor=\color{black!5}, %
  basicstyle=\footnotesize,%
}

\lstdefinestyle{customcpp}{
aboveskip=0.1in,
belowskip=0.1in,
abovecaptionskip=0.08in,
belowcaptionskip=0in,
captionpos=b,
xleftmargin=\parindent,
language=c++,
showstringspaces=false,
basicstyle={\linespread{0.9}\fontseries{sb}\footnotesize\ttfamily},
keywordstyle=\bfseries\color{keywordcolor},
commentstyle={\itshape\color{green!40!black}\fontsize{8}{8}\selectfont},
backgroundcolor=\color{white},
frame=single
}

	\begin{lstlisting}[style=customcpp]
  for (i = 0; i < N; i++) {
    //  launch a preemptible function 
    // and run it until the timeout 
    fn_launch(my_function, &fn_args[i], 
	&functions[i], timeout_us);
    // queue it for later execution if uncompleted 
    if (!fn_completed(&functions[i]))
       enqueue(run_queue, &functions[i]);
  }
    // round-robin scheduler 
  while (!empty(run_queue)) {
    // resume the function at the top of the queue 
    f = dequeue(run_queue);
    fn_resume(f, timeout_us);
    // queue it for later execution if uncompleted 
    if (!fn_completed(f))
       enqueue(run_queue, f);
  }
\end{lstlisting}
\centering
\footnotesize
\caption{LibPreemptible example of a simple round-robin scheduler running N static user-level threads. }
\label{fig:preemption_example}
\end{figure}
\vspace*{-0.2in}




\vspace*{0.15in}
\pagestyle{plain}

\section{Evaluation}
\label{sec:evaluation}

\LibName can be used by any datacenter applications written in C/C++. 
We deploy the system on an Intel Xeon Scalable Processor codenamed Sapphire Rapids with UINTR support (the network stack is DPDK or kernel TCP, and all machines run on Linux kernel version 5.15.0-rc1+ with turbo-boost disabled and fixed frequency scaling at 1.7GHz). The server has 56 CPUs, 112 hyperthreads and 2 sockets. Hyperthreading is enabled unless noted. For older CPUs, \LibName will fall back to standard interrupts. Below we answer the following questions: 
\begin{itemize}[leftmargin=*]
  \item Does LibPreemptible outperform prior preemptible scheduling techniques in terms of latency and throughput? 
  \item Can LibPreemptible be deployed with third-party applications out-of-the-box? 
  \item What is the overhead of LibPreemptible? 
  \item Does adaptively setting the preemption quantum optimize performance and latency in a colocation environment? 
\end{itemize}

\begin{figure*}
  \centering
    \centering
    \begin{subfigure}{0.68\textwidth}

      \centering
      \includegraphics[scale=0.64,viewport=80 20 500 160]{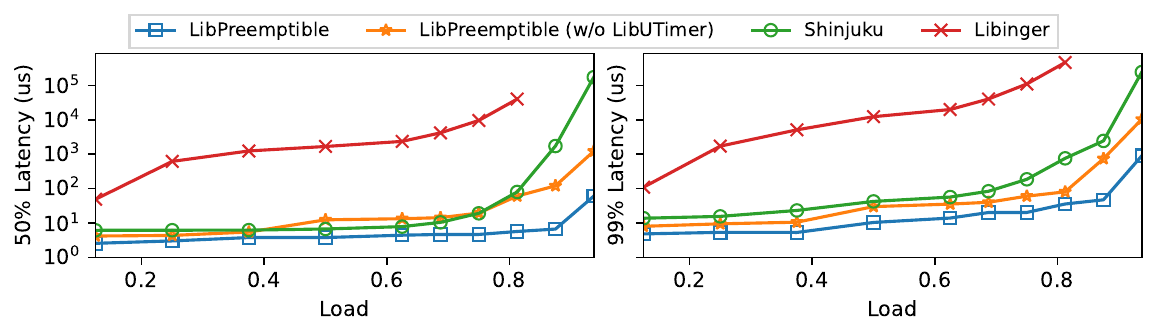}
      \label{fig:shinj_mediam}
    \end{subfigure}
     \hspace{-0.04in}
    \begin{subfigure}{0.3\textwidth}
      \centering
          
    \footnotesize{
    \begin{tabular}{cccc}
        \hline\hline
        Workload & A(1/2) & B & C \\
        \hline
        Shinjuku & 0.9 / 0.50 & 0.70 & 0.51\\
        Libinger & 0.35 / 0.23 & 0.12 & NA \\
        LibPreemptible & 1.1 / 0.75 & 0.78 & 0.68 \\
    \hline\hline
    \end{tabular} 
    \vspace{0.18in}
    }

    \end{subfigure}
    \caption{Left: Median and tail latency under different throughput for a bimodal workload ($0.5\%$ 500 us, $99.5\%$ 0.5 us). X axis is normalized with respect to the max load. Right: Tail latency bounded throughput (MRPS).  }
    \vspace{-0.02in}
    \label{fig:shinj}
\end{figure*}

\subsection{Performance Comparison}

We first compare LibPreemptible to Shinjuku~\cite{Shinjuku} and Libinger~\cite{libinger}, the most recent related work for preemption-based scheduling system. 
We first compare the three systems on synthetic loads, then on real workload. The synthetic workload is a server application where requests perform dummy work that we can control to emulate any target distribution of service times. 

We experiment with several types of workloads with poisson arrival rates. These distributions are selected to match workloads found in object stores and databases that mix simple GET/PUT requests with complex range or relational queries~\cite{Lim2014,Ousterhout2019}. 
\textbf{A. Heavy tailed;} 1) a bimodal workload with 99.5\% 0.5us and 0.5\% 500us requests, and 2) a bimodal workload with 99.5\% 5us and 0.5\% 500us requests. 
\textbf{B. Lighter tailed;} an exponential workload with mean request time of 5us. \textbf{C. Dynamic;} a workload with first half as heavy tailed (A1) and second half as lighter tailed (B), representing a distribution shift in client request patterns.

We experimented with different worker numbers. To ensure a fair comparison and the benefit even at the cost of one timer core, the experiments present 1 network thread, 5 worker threads for Shinjuku and Libinger, and 1 network thread, 4 worker threads (+1 timer thread) for LibPreemptible, running for 2 minutes.

Figure~\ref{fig:shinj} shows the comparison of  median (left) and tail (right) latency and throughput for the three systems. Under high load, the median and tail latencies with LibPreemptible are $\sim 10\times$ better than Shinjuku. The maximum throughput is measured by bounding 99\% tail latency by 200x the average latency in a stable system. Shinjuku needs to do careful profiling to select the right time quanta to achieve the desired throughput. LibPreemptible achieves better throughputs while dynamically adjusting the quanta under different workloads, as shown in Figure~\ref{fig:shinj} (right), 22\% higher than Shinjuku under workload A1, and 33\% higher than Shinjuku under C.

To separate the benefit from the new hardware, we disabled UINTR in LibUtimer (orange line), which makes the library run on ordinary timed interrupts. As we will show later, because the interrupt-based timer's granularity is much worse, and the overhead of timer delivery is also larger, the tail latency under higher load becomes worse by more than 5x. 

\begin{figure}[t]
  \centering
  \includegraphics[width=0.96\columnwidth]{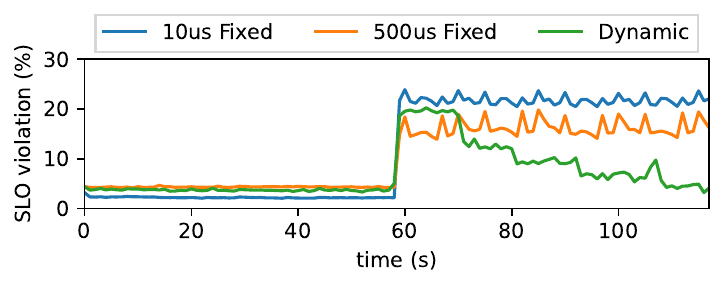}
  \vspace*{-0.1in}
	\caption{Dynamic LibUtimer is better at adapting to a distribution shift (Workload C), resulting in much fewer SLO violations. The latter half of the workload exhibits a higher average service time so the SLO violation rate is higher under a static policy. }
  \centering \label{fig:ex1-dynamic}
  \vspace*{-0.05in}
  \end{figure}

To evaluate the benefit of policy and deadline abstraction, Figure~\ref{fig:ex1-dynamic} shows how adaptive time quanta reduce SLO violations (as 50us) in workload C. The analysis to trigger the change in time quantum is only called every 10 sec and it is off the critical path, so it does not hurt the tail latency. Under lower load and low dispersion in service time, the time quantum is set to a higher value, consuming fewer CPU cycles for preemption. 

\subsection{LibPreemptible Analysis}
\label{sec:eval_analysis}

\begin{figure}[t]
  \centering
  \includegraphics[width=8.4cm]{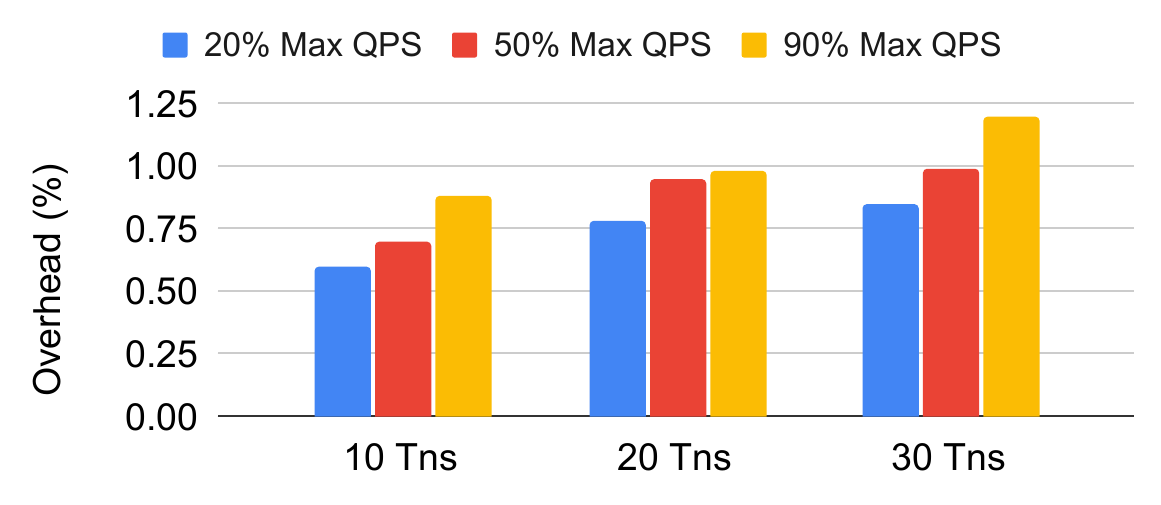}
	\caption{End-to-end percentage overhead impact on 99\% tail latency (Overhead \%= $\frac{p99_{libfunc} - p99}{p99}$) across different loads with different numbers of user-level threads per kernel thread ($T_n$) compared to no preemption.}
 \vspace*{-0.1in}
  \centering \label{fig:overhead}
  \end{figure}
  
  \noindent{\bf LibPreemptible deployment overhead: }We now measure the overhead of LibPreemptible for a simple gRPC server that uses no preemption by default. There are different threading models and configurations provided by the gRPC interface. 
We choose a thread pool threading model with blocking to ensure a low-latency baseline. LibPreemptible can also be incorporated on other threading models like Single Process Event-Driven (SPED)~\cite{SPED} architecture which operates on asynchronous ready sockets. 

We use a modified version of the open loop wrk2~\cite{wrk2} workload generator to generate requests with exponential service time. 
We measure the latency distributions at different QPS levels, with different numbers of user-level threads and kernel-level threads. We first fix the number of worker threads and measure the overhead of LibPreemptible with different numbers of user-level threads and user contexts at 20\% - 90\% of peak throughput, in terms of percentage degradation on p99 latency. 
As shown in the Figure~\ref{fig:overhead}, when the QPS is 20\% - 80\% of max load, the latency difference caused by LibPreemptible is negligible (below 1\%) as the number of thread contexts increases, showing that hardware-assisted interrupt 
delivery scales well on larger systems.

We also study how the tail latency behaves under different loads. 
When the load is around 89\% of max, we observe around 1.2\% tail latency overhead, due to the higher number of requests per thread that are preempted per unit of time. Overhead increases sublinearly for higher loads and is minimal. 
The deployment overhead for LibPreemptible is also minimal: in our real workload setups, we deployed LibPreemptible in just 1 week (Sec~\ref{sec:coloc}), which also included familiarizing ourselves with MICA and Zlib’s original threading library. All the changes needed to integrate LibPreemptible are at user-level, while for Libinger and Shinjuku, we need to either customize the kernel or customize the glibc library. The code needed is only 3\% of the original application code (excluding library code) to integrate LibPreemptible. The following tables show the additional time and code needed for integration (Table~\ref{tab:time-spent} and ~\ref{tab:code-spent}).

\begin{table}[!htp]\centering
\footnotesize
\begin{tabular}{lrrrr}\toprule
Time needed &MICA &Zlib &RPC \\\midrule
LibPreemptible &4 hours &3 hours &7 hours \\
Shinjuku &NA &NA &11 hours \\
Libinger &9 hours &8 hours &12 hours \\
\bottomrule
\end{tabular}
\caption{Time spent on integration by a researcher with no prior knowledge of either of the three frameworks. }\label{tab:time-spent}
\end{table}

\begin{table}[!htp]\centering
\footnotesize
\begin{tabular}{lrrr}\toprule
Code needed &MICA/Zlib &RPC \\\midrule
LibPreemptible &3\% &4\% \\
Libinger &NA &7\% \\
\bottomrule
\end{tabular}
\caption{Additional code percentage. Since Shinjuku has a high percentage of kernel code we did not include it. }\label{tab:code-spent}
\end{table}

\begin{table}[ht]
  \hspace{-0.3in}
  \centering
  \footnotesize
  
  \begin{tabular}{c|ccccc}
  \hline 
  & avg (us) & min (us)  & std (us) & rate (msg/s)  \\ \hline
  signal     & 15.325  & 3.584     & 3.478   & 63493 \\
  mq         & 10.468  & 8.960   & 2.017  & 95093 \\
  pipe       & 17.761  & 10.240    & 4.304  & 56151 \\
  eventFD    & 29.688  & 2.816   & 13.612  & 33629 \\ 
  uintrFd    & 0.734  & 0.512   & 0.698  & 857009 \\ 
  uintrFd (blocked)  & 2.393  & 2.048  & 0.212  & 409734 \\ 
  \hline
  \end{tabular}
  \caption{Overhead of different IPC mechanisms. }
  \label{tab:ipcoverhead}
  \vspace{-0.12in}
\end{table}
\noindent{\bf{Microbenchmarking: }}We now compare the interrupt delivery overhead for LibPreemptible to other IPC or event nofication mechanisms in Table~\ref{tab:ipcoverhead}. We adapt the microbenchmark suite~\cite{IpcBench}, 
and use 1M ping-pong IPC notifications with message size=1B.  
The user interrupt (averaged between blocked and running) has $10\times$ better average latency compared to the fastest IPC mechanism (message queue). This enables us to deliver fine-grained interrupts to user code. 
Moreover, the delivery overhead has smaller variance, even when the receiver is blocked. 
\begin{figure}[t]
  \centering
  \includegraphics[width=6.5cm]{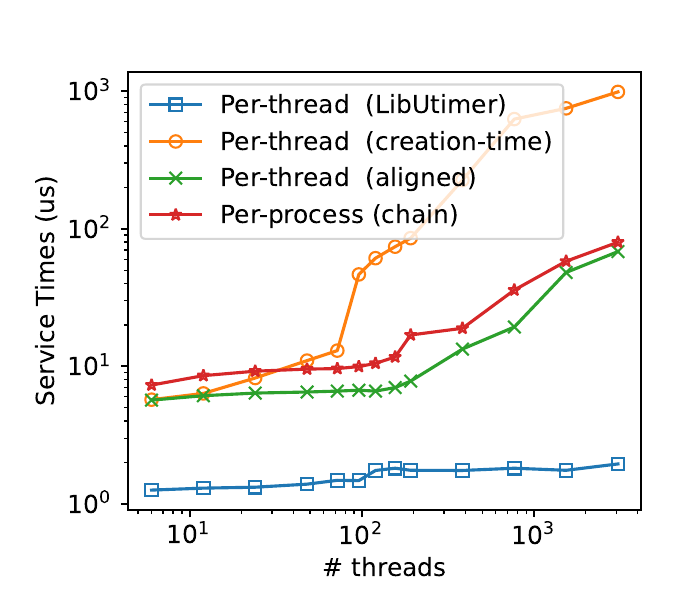}
  \vspace{-0.02in}
  \caption{Scalability of timer delivery overhead. }
  \centering \label{fig:scalability}
  \vspace{-0.08in}
  \end{figure}

  \begin{figure}[t]
  \centering
  \includegraphics[width=0.98\columnwidth]{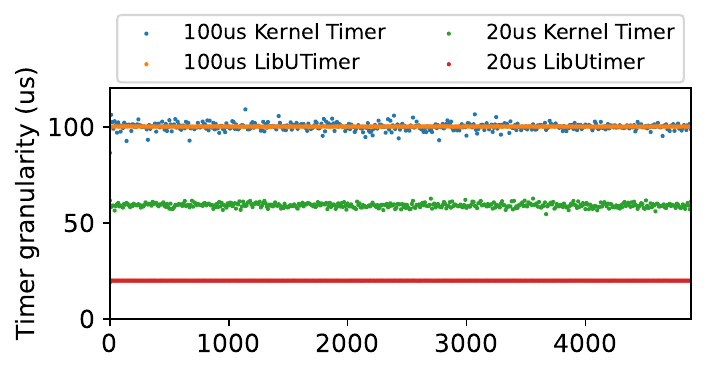}
  \vspace*{-0.1in}
	\caption{Precision of LibUtimer. X axis shows number of samples.}
  \centering \label{fig:user-timer}
  \vspace*{-0.15in}
  \end{figure}

  \noindent{\bf{Scalability and overhead of LibUtimer: }} We consider two approaches to implementing preemption timers in LibUtimer: per-thread and per-process. With a per-thread timer, every thread has its own OS timer. With a per-process timer, all threads in the process share the same OS timer: one thread receives the signal and forwards it to other threads.
  Figure~\ref{fig:scalability} shows the timer interrupt delivery overhead with 1000 interrupts on multiple threads, with interval between timer interrupt as 100us. %

  \begin{itemize}[leftmargin=*]
	   \item {\bf Per-thread (creation-time): }A na{\"i}ve implementation of a per-thread timer in which the timer of each thread is set independently (e.g., on thread creation) does not scale well on systems with large core counts. In Linux, calling a signal handler involves taking a lock in the kernel, thus causing lock contention when multiple signals are issued at the same time. Figure~\ref{fig:scalability} shows that the superlinearity is likely because of the lock contention, taking as much as 100 us for large core counts.
    \item {\bf Per-thread (aligned): }The timer interrupts across the different threads are explicitly aligned to reduce contention in the kernel lock. This approach significantly reduces the timer interrupt overhead, especially for higher thread counts, by almost 10$\times$ with 32 threads. However, it comes at a cost - the precision of the timer is impeded due to the delay of the issued interrupt. 
    \item {\bf Per-process (chain): }Shiina et al. proposed a new optimization to per-process timers, called ``chained signals''~\cite{Shiina2021}. The thread receiving interrupts handles the signal and then issues a signal to at most one other thread. 
    \item {\bf Per-thread (user-timer/LibUtimer): } LibUtimer achieves the best scalability across thread counts. 
  \end{itemize}

  Moreover, compared with the scalability limitations of earlier approach used by Shinjuku (since the APIC supports only a limited number of logical processors), LibUtimer are able to scale to more tenants using more logical processors by design. 
  
  \noindent{\bf{LibUtimer precision and power cost: }} Finally, we study and justify the timer thread overhead: 
  As a justification of dedicating a core for timer threads, we measure the cost is about 1.2 Watts for the first core because UMWAIT can save energy for busy polling. With each additional core, the power cost is minimal. Besides, we also use \texttt{stress-ng}~\cite{stress-ng} to inject some contentions, and even with kernel activities (IRQ-affinity, TLB shootdowns, overcommitment), we observed that the timer preciseness will not be significantly impacted. 
  In Figure~\ref{fig:user-timer}, we measure the jittering of LibUtimer timer with background activities, when our target quanta is 100us and 20us. It shows the time between setting up a periodic kernel timer vs LibUtimer with 26 threads and visualizes the time (y-axis) between calls to the handler for 5000 consecutive samples (x-axis). Kernel timer's granularity cannot go down to $20us$ (which is why we see a line around $60us$) and shows high variations, and LibUtimer can consistently give precise timer (average relative error for timer delivery is around 1\% for 5000 samples).

\subsection{LibPreemptible in Real Workload Scenarios}\label{sec:coloc} 

To improve resource management and enable efficient CPU utilization, latency critical (LC) jobs can be time-sharing CPU resources with other applications, often best effort (BE), and use preemption to reclaim resources. This reclaimation must happen adaptively in micro-second scale to ensure low tail latency when needed. Achieving this in practice is challenging, especially for the jobs with $\mu$s- and sub-$\mu$s- SLO requirements, due to relatively large overhead of preemption in current systems. With LibPreemptible, this overhead is dramatically reduced; this can allow low-latency LC jobs to meet their SLO even when aggressively sharing resources with BE tasks. 

\subsubsection{Experiment Setup}

We use the MICA~\cite{Lim2014} KVS system as the LC work in this experiment due to its sub-$\mu$s scale execution time for small requests. As the co-located BE processing, we use data compression based on zlib~\cite{zlib}. We run MICA under 5/95 SET/GET request distribution with the skewness at 0.99. We use the default zipfian generator from the original MICA work. This yields a median request processing time of 1us. As the compression workload, we run zlib engines against 25~kB of raw data where median latency is 100~us. Table~\ref{tab:co_location_workload} summarizes the configuration of the datasets used for both workloads, and shows the median and tail request latencies when running them individually, without co-location, and on a single core.

\begin{table}[]
  \centering
  \footnotesize
  \begin{tabular}{ccccc}
    \hline\hline
	  \multicolumn{1}{c|}{{\bf App}}  & \multicolumn{1}{c|}{{\bf Params}}      & \multicolumn{1}{c}{{\begin{tabular}[c]{@{}c@{}}\bf Latency\\ {\bf 50\%, 99\%} \end{tabular}}} &  \\ \cline{1-4}
 \multicolumn{1}{c|}{{MICA}} & \multicolumn{1}{c|}{\begin{tabular}[c]{@{}c@{}}EREW mode\\ \{k, v\} = \{16, 64\}B\\ SET/GET = 5/95\\ zipf w/ skewness 0.99\end{tabular}} & \multicolumn{1}{c}{1us, 7 us}  & \\  \hline
 \multicolumn{1}{c|}{{zlib}} & \multicolumn{1}{c|}{data size = 25 kB}    & \multicolumn{1}{c}{100us, 250 us}                                                                                  &  \\
 \hline\hline
 \end{tabular}
 \caption{Workloads for the co-location experiment. }
 \label{tab:co_location_workload}
 \vspace{-0.12in}
\end{table}

The request generator issues uniformly distributed BE and LC requests with 2\% and 98\% rates respectively with both constant and bursty QPS.
The dispatch thread is connected (over $dispatch\_queue$) to worker threads via our user-space request scheduler based on LibPreemptible. We showcase two different scheduling policies, to demonstrate how adaptability in the time quantum helps 
with achieving the best of both worlds for LC's SLO and BE's throughput. 

\begin{figure*}
  \centering
  \begin{subfigure}{0.46\textwidth}
    \centering
    \includegraphics[width=\linewidth]{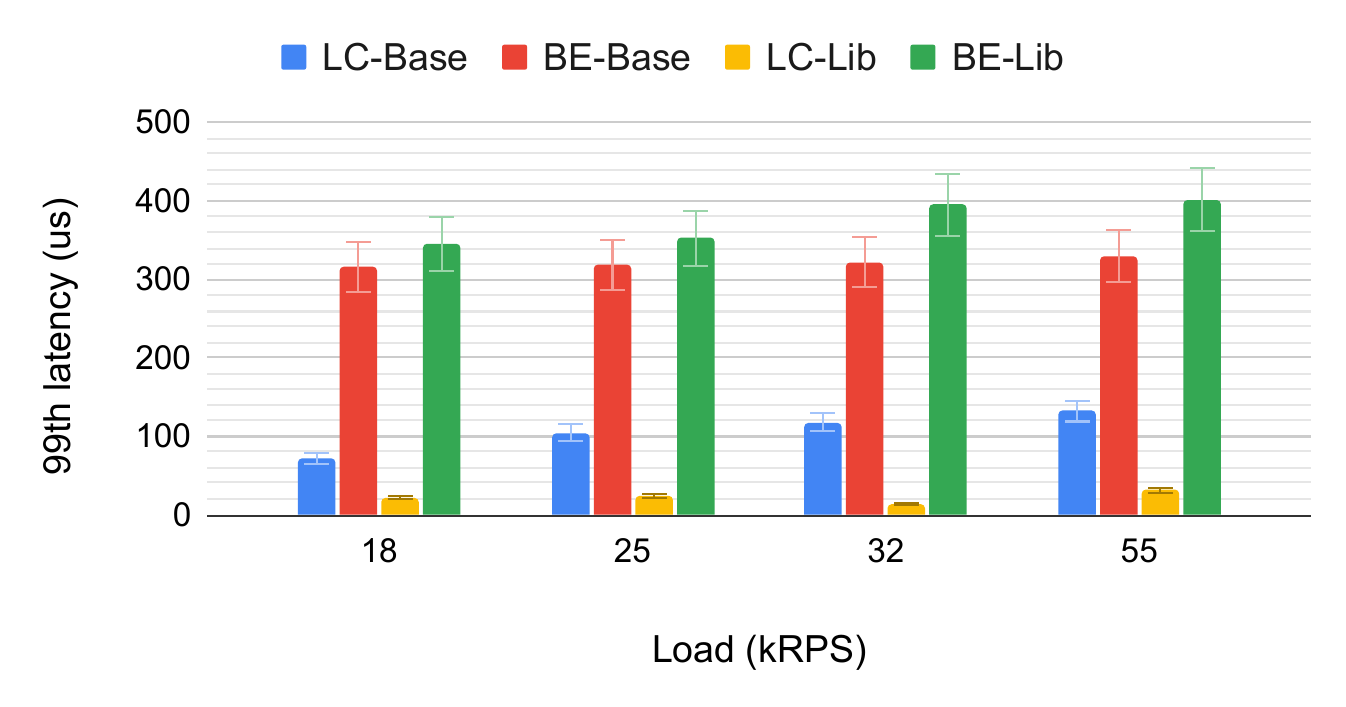}
    \label{fig:co_loc_1_a}
  \end{subfigure}
  \qquad
  \begin{subfigure}{0.46\textwidth}
    \centering
    \includegraphics[width=\linewidth]{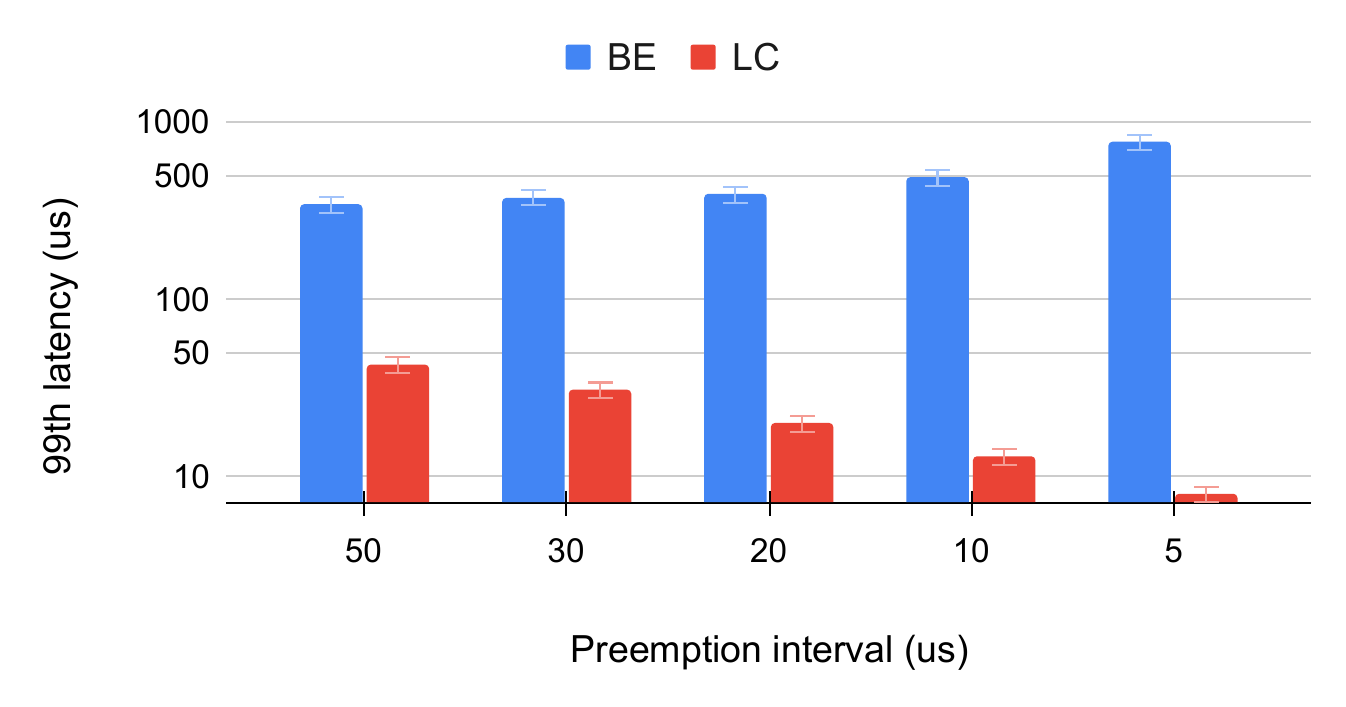}
    \label{fig:co_loc_1_b}
  \end{subfigure}
  \caption{Tail latency of co-located LC and BE jobs with LibPreemptible based preemptive scheduler with fixed time quantum of 30us (left) and with variable time quantum over fixed QPS of 55kRPS (right).}
  \label{fig:co_loc_1}
\end{figure*}

\subsubsection{Scheduling policy \#1, \textbf{FCFS with preemption, constant preemption interval}} Here the scheduler picks up the first request from the $dispatch\_queue$ and runs it in the worker thread with the static and constant time quantum of 30~us. If the request runs for a longer time period, it gets preempted and pushed into the $long\_queue$. Then the scheduler picks up the next element from the $dispatch\_queue$ thereby giving preemptive priority to shorter jobs. If the $dispatch\_queue$ is empty, the scheduler resumes previously preempted requests from the $long\_queue$ via $fn\_resume()$ calls. The results of the experiment are shown in Figure~\ref{fig:co_loc_1} (left).

As Figure~\ref{fig:co_loc_1} (left) shows, our preemptive scheduler brings the 99th tail latency of the LE job (LC-Lib) down $3.2\times - 4.4\times$ times compared to non-preemptive execution (LC-Base). With the preemption interval of 30~us, the scheduler brings the tail latency of MICA KVS requests down to 33~us under 55~kRPS. When the preemption interval is set to 5~us (Figure~\ref{fig:co_loc_1}, right), the scheduler brings the 99th tail latency of MICA requests down to 8~us, which is $18.5\times$ lower than non-preemptive execution under the same load and co-location environment. However, the preemption interval of 30~us results in 30\% increase of the BE job latency, and the overhead reaches $2.2\times$ when using very low intervals of 5~us. 

\subsubsection{Scheduling policy \#2, \textbf{FCFS with preemption, dynamic preemption interval}} One of the key benefits of the LibPreemptible API is the ability to dynamically set the preemption interval on a per-request basis. The ability to change the time quantum in runtime allows us to build an adaptive scheduling policy that would adjust preemption in interval depending on the current request rate. In this experiment, we modify the previous policy by adding two new components in the system: the QPS monitor, and the controller of the preemption interval.

\begin{figure*}[t]
  \centering
  \begin{subfigure}{.34\textwidth}
    \centering
    \includegraphics[width=\linewidth]{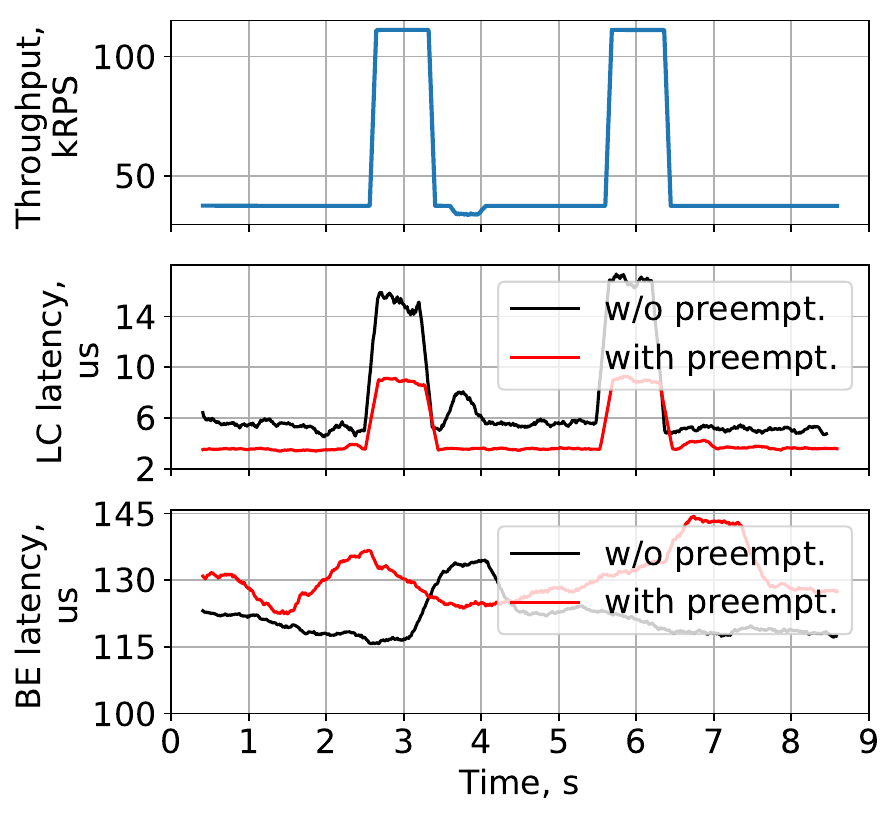}
    \label{fig:co_loc_1_a}
  \end{subfigure}
  \hspace{-0.5\baselineskip}
  \begin{subfigure}{.315\textwidth}
    \centering
    \includegraphics[width=\linewidth]{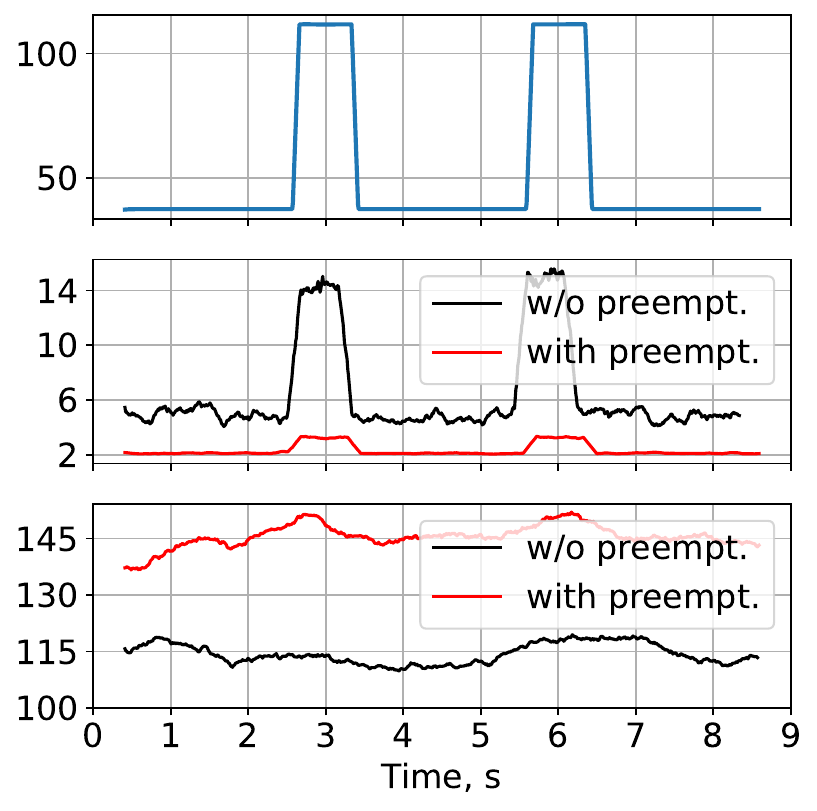}
    \label{fig:co_loc_1_b}
  \end{subfigure}
  \hspace{-0.5\baselineskip}
  \begin{subfigure}{.315\textwidth}
    \centering
    \includegraphics[width=\linewidth]{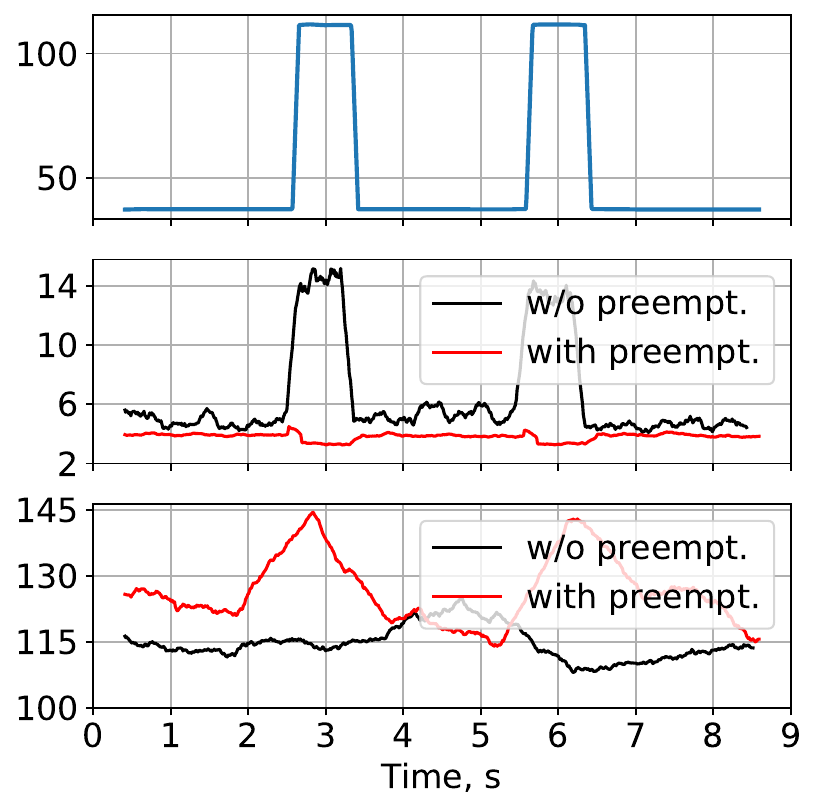}
    \label{fig:co_loc_1_b}
  \end{subfigure}
  \caption{Average latency of LC and BE jobs over time with a constant preemption interval of 50~us (left), 10us (middle), and with the dynamic policy (right); the top plot shows measured QPS of the bursty load in the dispatch thread, the middle and the bottom - average latency of LC and BE jobs respectively over the 10~s timeframe.}
  \label{fig:co_loc_2}
\end{figure*}

According to this policy, the scheduler monitors the current QPS rate of incoming requests and sets the preemption interval in the worker threads according to the load. We test the policy with a spiky load generator that periodically issues bursty traffic (Figure~\ref{fig:co_loc_2}). We allow the controller to set the preemption interval to values between 10 and 50~us, and our workload QPS changes from 40 to 110~kRPS.

Figure~\ref{fig:co_loc_2} (left) shows that a preemption interval of 50~us reduces the average latency of the LC job from 14~us to 10~us, under the QPS spikes. It does not affect the latency of the co-located BE job significantly when load is low, and only marginally during spikes. When using lower preemption intervals of 10~us (middle), we achieve a reduction of the LC average latency down to 3~us -- $5\times$ lower than when running without LibPreemptible, however the latency penalty for the BE job is relatively higher. While complete avoidance of the latency penalty of the BE job is impossible due to preemptions, the overhead can be reduced when the load is low by setting the preemption interval to a higher value during these periods. Figure~\ref{fig:co_loc_2} (right) shows performance with our adaptive time quantum scheduler. The average latency of the LC job remain low during the time of the experiment, while the negative impact on the BE jobs during periods of low load is minimized. These results justify the need for adaptive, low-overhead preemption when the workload is spiky, and showcase the corresponding scheduler that can be build with LibPreemptible.

\section{Related Work}
\label{sec:related_work}
\pagestyle{plain}

We now discuss prior proposals for scheduling for reducing tail latency and improving CPU efficiency, and quantify our differences.

{\bf Dataplane Operating Systems:}
A dataplane OS improves throughput and/or latency by separating its control and data planes. Initially, the systems that target microsecond-level workloads, such as Chronos~\cite{Kapoor2012}, IX~\cite{IX}, and Arrakis~\cite{Peter2015}, choose to statically pin threads to cores, and offload load balancing to the NIC hardware, thus eliminating the associated scheduling overheads.
This approach works well for homogeneous workloads, however, it can introduce higher latencies for complex workloads with varying execution times, e.g., key-value stores~\cite{Lim2014, Enberg2019}.
ZygOS~\cite{Prekas2017} shows that load-balancing through work stealing among the pinned threads is necessary even at these timescales.
Shinjuku~\cite{Shinjuku} demonstrates that it is feasible to implement more complex scheduling policies without significant overheads by re-purposing low-overhead preemption mechanisms normally used by VMs. \textit{These approaches require significant changes to the OS, and typically support specific types of applications, which are not scalable and do not coexist with common cloud system software stacks. LibPreemptible performs better than Shinjuku while operating \textit{safely} on top of the Linux kernel. Our work will complement existing dataplane OSes and hardware offloading techniques. 
}

{\bf User-level Threading:}
There is a significant body of work on co-operative userspace thread libraries starting from Adya et al.~\cite{Adya2002} who proposes the concept of userspace threading with automatic stack management.
Capriccio~\cite{vonBehren2003} introduces optimizations to improve scalability and scheduling, optimizing stack allocation while identifying some of the pitfalls of non-preemptive scheduling.
Inspired by scheduler activations~\cite{Anderson1992}, Arachne~\cite{Qin2018} makes userspace threading core-aware and  thread creation faster. Psyche~\cite{10.1145/121133.344329} introduces first-class user-level threads that delivers software interrupts. 
Commercial languages and libraries, such as Go~\cite{Golang}, folly::Fibers~\cite{folly}, Boost fibers, uThreads~\cite{uthread}, and C++ coroutines~\cite{Ccoroutine}), have adopted many of these optimizations.

Libturquoise~\cite{libinger} is the first attempt to create a general-purpose user-level threading library using regular timer interrupts as the preemption mechanism, and mainly focusing on handling shared state and non-reentrant code.
Shiina et al.~\cite{Shiina2021} improves upon libturquoise via chained per-process timers and general-purpose, kernel-level thread switching.
\textit{\LibName leverages hardware support to minimize the preemption overhead of user interrupts, and facilitates the development of custom user-level scheduling policies.} %

\begin{figure}[htb]
  \centering
  \includegraphics[width=\linewidth]{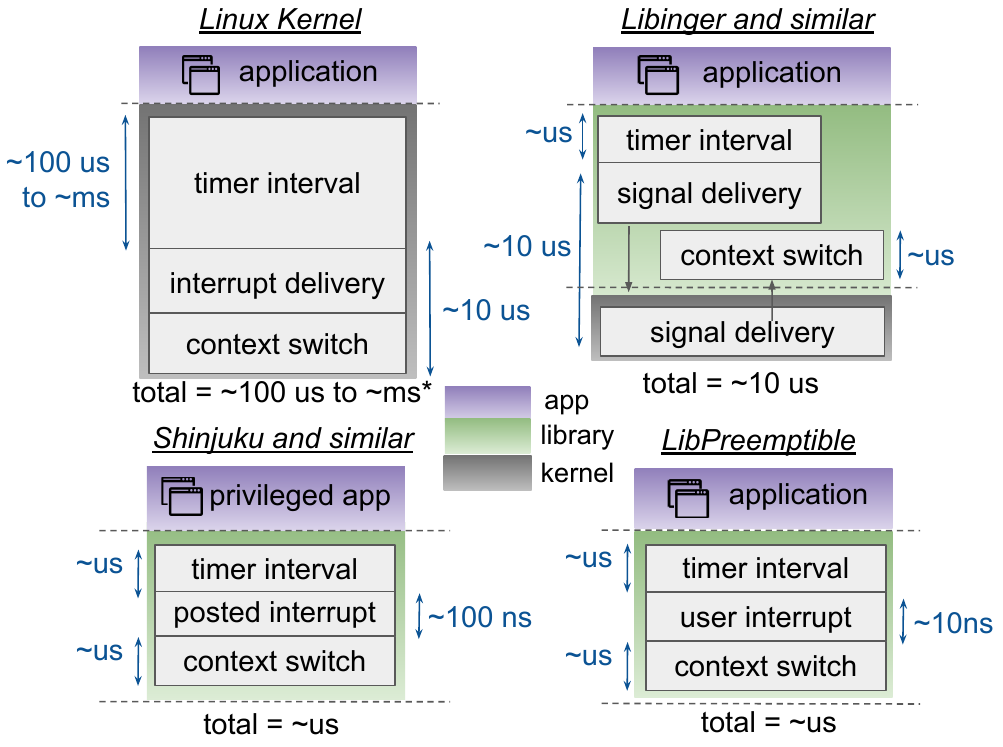}
	\caption{Comparison of \LibName to prior work. }
  \vspace{-0.15in}
  \label{fig:intro}
\end{figure}

\vspace{0.05in}
{\bf Scheduling Policies:}
Based on the fact that First-Come-First-Serve (FCFS) scheduling has been shown~\cite{Wierman2012} to be tail-optimal for light-tailed homogeneous tasks, many older systems did hash-based load balancing on the network interface card (NIC) using receive side scaling (RSS) and running requests to completion~\cite{Lim2014, Peter2015, IX}.
To handle imbalance between workers, newer systems enhance RSS to take into account end-host load (RSS++~\cite{Barbette2019}, eRSS~\cite{Rucker2019}), employ work-stealing (ZygOS~\cite{Prekas2017}, Shenango~\cite{Ousterhout2019}, Caladan~\cite{Fried2020}, BWS~\cite{Ding2012}, Elfen~\cite{ELfen}, Li at al.~\cite{Li2016}), or use techniques, such as join-idle-Queue~\cite{Lu2011} or join-bounded-shortest-queue~\cite{Kogias2019}.
To accommodate highly-variable workloads, Shinjuku~\cite{Shinjuku} takes a different approach by implementing centralized preemptive scheduling in a dedicated core, while Bertogna et al.~\cite{Bertogna2011} attempt to find the optimal preemption points.
Persephone~\cite{Demoulin2021} leverages application-specific knowledge to reserve cores for short requests and avoid preemption altogether.
Other proposals employ custom hardware with centralized scheduling (Mind the Gap~\cite{Humphries2021}, nanoPU~\cite{Ibanez2021}, RPCValet~\cite{Daglis2019}), priority queues (ExpressLane~\cite{ExpressLane}), or fast context switching~\cite{ContextSwitch}.

Recently, McClure et al.~\cite{McClure2022} show that the trade-offs between load balancing and core allocation are not easy to navigate even without considering request reordering and preemptive policies. It has generated a renewed interest in customizable and application-specific scheduling.  Zhao et al.~\cite{zhao2022altocumulus}
focus on scalable, adaptive, SLO-aware scheduling systems by combining the effectiveness of SLO-aware migration policy and a hardware-based mechanism for short-lived RPCs.
SKQ~\cite{Zhao2021} makes event scheduling on top of the Linux kernel configurable.
Ford et al.~\cite{Ford1996}, Slite~\cite{Gadepalli2020}, and ghOSt~\cite{Humphries2021} offload kernel thread scheduling decisions to userspace while Syrup~\cite{Kaffes2021} allows users to deploy custom scheduling functions throughout the stack safely.
\textit{\LibName does not require additional hardware, simplifies the specification and deployment of custom thread scheduling policies, providing complete flexibility in how preemption is employed. }

\if\sensitive0


\fi
Finally, \LibName is orthogonal to and can benefit existing work which dynamically allocates CPU cores like~\cite{Fried2020}, to schedulers that focus on microsecond-scale reallocation of resources, like Caladan~\cite{Fried2020} and Shenango~\cite{Ousterhout2019}, and to systems which use kernel bypass techniques, like DPDK~\cite{DPDK}. Figure~\ref{fig:intro} shows how LibPreemptible differs from prior work. %

\section{Discussion}
\pagestyle{plain}

\subsection{Native User Interrupts Security Discussion}
The security model for native User Interrupts is designed with a focus on trusted and cooperating processes. Note that the model allows any sender with access to \texttt{uintr\_fd} to generate the associated interrupt vector for the receiver task that created the \texttt{fd}. This could potentially lead to issues with untrusted processes launching a Denial of Service attack. 

The potential for a DoS attack by generating a storm of user interrupts is a valid concern. The fact that a user interrupt handler is invoked with interrupts disabled, but upon execution of \texttt{uiret}, interrupts get enabled by the hardware, could indeed lead to the handler being invoked before normal execution can resume.

LibPreemptible only allows timer threads to send preemption, which is under the same security domain. The quantum is also controlled by trusted application process. So LibPreemptible reduce the attack surface compared with native UINTR.




\subsection{Attack Surface under LibPreemptible}


Security of interrupt-handling is critical, especially in real cloud environments. Shinjuku uses inter-process interrupts (IPIs) directly, rather than Linux signals for preemption to reduce overhead, because it executes in privilege ring 0. To minimize the preemption overhead, it maps the local APIC for each application core to the same address space as the dispatcher. Because the APIC is directly mapped to ring 3 in Shinjuku, both the Shinjuku runtime and application must be trusted. Direct APIC access enables buggy or malicious code to flood the system with IPIs, creating a DoS attack vector against all cores. LibPreemptible avoids this exposure through the use of User Interrupts. While User Interrupts may be sent directly without kernel intervention, allowing low-latency signaling, they may only be sent to targets configured in the kernel-maintained User Interrupt Target Table (UITT). In LibPreemptible, the only configured vectors are between the timer cores and worker applications. Buggy runtime code can at most degrade application performance within a single runtime.





\subsection{Other Use Cases and Future Work}
Other use cases for Libpreemptible include traffic shaping~\cite{Carousel}, scheduling in 5G~\cite{Concordia}, and real-time DNN serving~\cite{osdi2022reef}. 
The accuracy of these timed actions is crucial to the performance of these real-time applications. Concurrent DNN serving with lightweight micro-second scale preemption on CPU can also improve the throughput and latency. \LibName can provide a precise and lightweight solution to these use cases. 

Additionally, while our evaluation uses a dedicated core to generate timer events using the UINTR capability, it is relatively easy to offload this capability directly to hardware. 
In fact, hardware vendors are exploring supporting this type of capability using a dedicated hardware timer that can deliver an interrupt directly to the application~\cite{Patent}. This hardware offloading approach will improve performance/watt at the cost of area of the chip. 

\section{Summary}
We propose LibPreemptible, a hardware-assisted library for user-level scheduling. Our primitives achieve flexibility, good performance, and scalability at the same time.  LibPreemptible leverages user-level interrupts, and its abstraction delegates scheduling decisions to applications and provides an interface that supports dynamic policies. 
Compared with prior systems it showed significantly better performance and scalability. LibPreemptible is compatible across application types and requires no changes to the existing OS kernels, making it easy to be deployed in cloud settings. This work can both highlight the benefits of hardware-assisted user-level scheduling and motivate follow-up work that improves the generality of dynamic application-aware scheduling policies.

\section{Acknowledgement}
We sincerely thank Qizhe Cai, Chris De Sa, Sihang Liu, Kevin Tang, Yang Zhou, Sol Boucher, Tom Anderson, Ken Birman, Mengjia Yan for their feedback. We thank the colleagues in SSR group in Intel Labs (especially Rajesh Sankaran and James Tsai) for the support. This work was supported in part by NSF CAREER Award CCF-1846046, and an Intel Research Award. 
\balance

\clearpage

\bibliographystyle{plain}
\bibliography{refs}

\end{document}